\documentclass[aps, prl, twocolumn, superscriptaddress, showpacs]{revtex4}
\usepackage{graphicx}
\usepackage{dcolumn}
\usepackage{bm}
\usepackage{natbib}
\usepackage{upgreek}
\usepackage{amsmath}
\usepackage{amssymb}

\begin{document}

\title{Universal Faraday rotation in HgTe wells with critical thickness}

\author{A. Shuvaev}
\author{V. Dziom}
\affiliation{Institute of Solid State Physics, Vienna University of
Technology, 1040 Vienna, Austria}
\author{Z. D. Kvon}
\author{N. N. Mikhailov}
\affiliation{Novosibirsk State University, Novosibirsk 630090, Russia}
\affiliation{Institute of Semiconductor Physics, Novosibirsk 630090, Russia}
\author{A. Pimenov}
\affiliation{Institute of Solid State Physics, Vienna University of
Technology, 1040 Vienna, Austria}

\begin{abstract}
The universal value of Faraday rotation angle close to the fine structure constant ($\alpha \approx 1/137$) is experimentally observed
in thin HgTe quantum wells with thickness on the border between trivial insulating and the topologically non-trivial Dirac phases. The quantized value of the Faraday angle remains robust in the broad range of magnetic fields and gate voltages. Dynamic Hall conductivity of the hole-like carriers extracted from the analysis of the transmission data shows theoretically predicted universal value of $\sigma_{xy}=e^2/h$ consistent with the doubly degenerate Dirac state. On shifting the Fermi level by the gate voltage the effective sign of the charge carriers changes from positive (holes) to negative (electrons). The electron-like part of the dynamic response does not show quantum plateaus and is well described within the classical Drude model.
\end{abstract}

\date{\today}

\maketitle

The strong spin-orbit coupling and an inverted band structure in mercury telluride makes this material to a nearly universal tool to probe novel physical effects with the film thickness being a tuning parameter~\cite{bernevig_science_2006, buttner_nphys_2011}. If the thickness of HgTe wells is below critical, $d<d_c\approx 6.3$~nm, the sequence of the conduction and valence bands is conventional and a trivial insulating state is realized.
For thicker films and in the bulk mercury telluride the inversion of valence and conduction bands leads to topologically non-trivial surface states~\cite{qi_prb_2008, hasan_rmp_2010}. This state is characterised by the locking of the electron spin and the electron momentum and they are topologically protected against non-magnetic impurity scattering.

If the thickness of HgTe well is equal to critical, the gap between valence and conduction bands disappears and a two-dimensional (2D) electron gas is formed with Dirac cone dispersion~\cite{bernevig_science_2006, buttner_nphys_2011}.
Close to the center of the Dirac cone the electron spin is not a good quantum number, but has to be replaced by pseudo-spin or helicity~\cite{neto_rmp_2009, qi_prb_2008}. Due to the particle-hole symmetry of these states, the quantum Hall effect becomes shifted by a half-integer and takes the form $\sigma_{xy}=\gamma(n+1/2)e^2/h$. In well-investigated case of graphene~\cite{novoselov_nature_2005, zhang_nat_2005} the states are fourfold degenerate, i.e. $\gamma=4$, as two Dirac cones are present in the Brillouin zone which are both doubly spin-degenerate.

Magneto-optics in the terahertz range has been proven to be an effective tool to investigate two-dimensional conducting states in several quantum systems, like graphene~\cite{crassee_nphys_2011, shimano_ncomm_2013, orlita_njp_2012}, Bi$_2$Se$_3$~\cite{schafgans_prb_2012, bordacs_prl_2013, wu_np_2013, olbrich_prl_2014}, and HgTe~\cite{shuvaev_prl_2011, hancock_prl_2011,kvon_jetpl_2011, zholudev_prb_2012, shuvaev_prb_2013, olbrich_prb_2013, zoth_prb_2014, dantscher_prb_2015}. Magneto-optical spectroscopy has the advantages of being contact-free and of directly accessing the effective mass $m_c$ via the cyclotron resonance $\Omega_c = eB/m_c$. Here $B$ is the external magnetic field.

In the dynamical regime the unusual character of the quantum Hall effect in systems with Dirac cones can be shown~\cite{tse_prb_2010, maciejko_prl_2010, tse_prl_2010, tkachov_prb_2011}  to lead to a universal values of the Faraday and Kerr rotation with $\theta_F=\alpha \sim 1/137$ and $\theta_K=\pi/2$, respectively. Such predictions have been recently confirmed experimentally in graphene~\cite{shimano_ncomm_2013}, where the Faraday angle is additionally doubled as two Dirac cones exist in the Brillouin zone. Very recently~\cite{okada_nc_2016, wu_arxiv_2016, dziom_arxiv_2016}, several groups announced the observation of the quantized Faraday and Kerr rotation from the surface states of various topological insulators. In our previous work ~\cite{dziom_arxiv_2016} the universal Faraday rotation has been observed on surface states in three-dimensional topological insulator, realised in thick strained HgTe film.

Compared to thick strained films with three-dimensional (3D) carriers, in HgTe wells with critical thickness a two-dimensional electron gas is realized. In this case there is no gap between the valence and conduction bands, resulting in Dirac dispersion of the 2D carriers. Due to double degeneracy of the Dirac cone the quantized dynamical Hall conductivity with $\gamma = 2$ may be expected.

In this manuscript we present the results of the terahertz experiments in HgTe quantum wells with critical thickness. Compared to graphene~\cite{shimano_ncomm_2013} here a single valley Dirac fermion system is realized revealing strong spin-orbit coupling. Critical sample thickness ensures 2D character of charge carriers which can be tuned using a transparent gate.

\begin{figure}[tbp]
\centerline{\includegraphics[width=0.75\columnwidth,clip]
{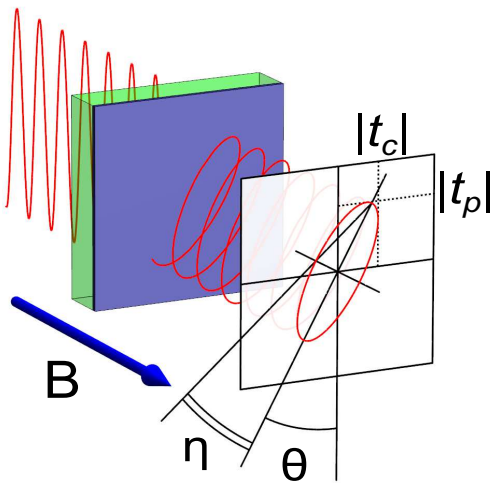}}
\caption{Schematic view of the magneto-optical experiment to measure the Faraday rotation $\theta$ and ellipticity $\eta$. The definitions of both angles are depicted in the output polarisation ellipse, assuming linear incident polarisation. The external magnetic field is applied in the Faraday geometry, i.e. $B \| \mathbf{k}$. Complex transmission in parallel, $t_p$, and in crossed, $t_c$, polarisers are measured which provide full description of the transmission matrix.}
\label{fig_exp}
\end{figure}

Mercury telluride quantum wells have been grown on (013) oriented GaAs substrates by molecular beam epitaxy as described elsewhere~\cite{kvon_ltp_2009}. The results on two samples with thickness close to critical are presented: sample \#1 with $d=6.3$~nm and sample \#2 with $d=6.6$~nm. The gate on both samples has been prepared \emph{ex-situ} using a mylar film with $d=6 \mu$m as an insulating barrier and a semi-transparent metallized film as a gate (Ti, $R=600 \Omega/ \Box$). In the experiment the gate conductivity is seen as magnetic field-independent and frequency independent contribution to $\sigma_{xx}$. No measurable effect of gate on the Hall conductivity has been observed which agrees well with low mobility of the gate carriers.

The experimental results in this work have been obtained using a Mach-Zehnder
interferometer operating at sub-millimeter wavelengths (0.1-1~THz)~\cite{volkov_infrared_1985}.  The interferometric arrangement has enabled to obtain the absolute values of complex transmissions through the sample in parallel and crossed polarisers geometries~\cite{shuvaev_sst_2012}. The experimental procedure is shown schematically in Fig.~\ref{fig_exp}. External magnetic fields up to 7~T have been applied using a superconducting magnet with polypropylene windows.
The transmission experiments have been done in the
Faraday geometry, i.e.  magnetic field is applied along the
propagation direction of the electromagnetic radiation. In total, four experimental parameters are measured and full characterization of the transmitted radiation is obtained including the polarization state. With this information, the Faraday rotation angle $\theta$ and
ellipticity $\eta$ can be obtained directly~\cite{shuvaev_sst_2012,shuvaev_prb_2013, dziom_2d_2016}.

Explicit equations to calculate the matrix of the conductivity from the measured transmission are given in Ref.~\cite{dziom_2d_2016}. In these calculations the effect of GaAs substrate and of Ti gate are taken into account exactly, i.e. pure HgTe conductivity is obtained. The measured Faraday rotation and ellipticity are still partly influenced by the properties of substrate and gate. Where appropriate, specific values of these angles will be given. The frequency of transmission experiment is chosen to minimize the influence of the substrate.

\begin{figure}[tbp]
\centerline{\includegraphics[width=0.95\columnwidth,clip]
{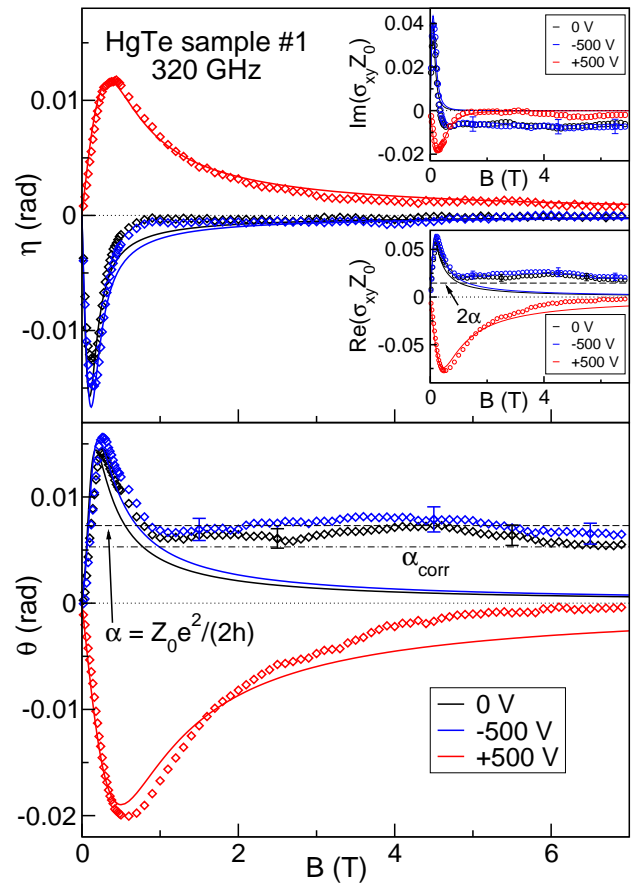}}
\caption{Magnetic field dependence of the Faraday rotation $\theta$ (lower
panel) and ellipticity $\eta$ (upper panel) for the sample \#1 for three characteristic gate
voltages. Experimental data are shown by solid symbols and the lines are fits
within the Drude model~\cite{tse_prb_2011, tkachov_prb_2011}.
Dashed line shows a "pure" universal value of Faraday rotation $\alpha \approx 1/137$~rad. Dash-dotted line gives the real value of the rotation $\alpha_{\mathrm{corr}}$ taking into account the influence of the substrate and gate, and assuming $\sigma_{xy}=e^2/h$.
The inset shows the off-diagonal conductivity
$\sigma_{xy}$ as directly obtained from the spectra using exact expressions for magneto-optical transmission~\cite{dziom_2d_2016}.}
\label{theta_eta_7T}
\end{figure}

The most important result of this paper is demonstrated in
Fig.~\ref{theta_eta_7T}. Here, the experimental Faraday rotation $\theta$
(lower panel) and ellipticity $\eta$ (upper panel) are shown for the sample \#1. The peaks in the data at around $B\sim 0.5$~T are the cyclotron resonances on
free charge carriers in our sample. The sign change of the Faraday
angle and ellipticity between negative and positive gate voltages corresponds to the transition from the hole-like to the
electron-like carriers, respectively.

We note that in high magnetic fields far above the cyclotron resonance~\cite{shuvaev_apl_2013}, classical Faraday rotation and ellipticity are
expected to fade out as $\theta \sim 1/B, \eta \sim 1/B^2$.
Remarkably, in Fig.~\ref{theta_eta_7T} the experimental values of the Faraday rotation for zero and negative
gate voltages saturates at fields above 1~T and stays constant within
the experimental accuracy up to the highest field in our experiment (7~T). We note that similar broad steps in the quantum Hall resistivity have been recently observed in HgTe wells and attributed to heavy holes valleys reservoir effects~\cite{kozlov_jetpl_2015}.
The step in Faraday rotation reveals a universal
value close to the fine structure constant $\alpha = \frac{Z_0}{2}\frac{e^2 }{h} $,
indicated in Fig.~\ref{theta_eta_7T} by dashed lines. Dash-dotted line gives the value which takes into account the properties of the substrate and gate exactly~\cite{dziom_2d_2016}, and assuming $\sigma_{xy}=e^2/h$, $\sigma_{xx}=0$. The difference between both values of Faraday rotation and the experimental data are within the uncertainties of the present work.
As discussed above, in HgTe with critical thickness a 2D electron gas with Dirac dispersion is realized with double degeneracy~\cite{bernevig_science_2006, buttner_nphys_2011}. Therefore, we attribute the observation of  $\theta \approx \alpha$ to doubly degenerate states each contributing by $\alpha/2$.

In order to demonstrate the discrepancy between the experimental data and
classical cyclotron resonance, the fits within Drude model~\cite{tse_prb_2011, tkachov_prb_2011}
are shown as solid lines in
Fig.~\ref{theta_eta_7T}. The Faraday rotation $\theta$ and the Faraday ellipticity $\eta$ of the electron-like carriers at
positive gate voltage are well fitted within the classical response (red lines and symbols).
The ellipticity at zero and negative gate voltages also follow the
classical Drude model quite well. Remarkably, the experimental Faraday rotation in this region of
the gate voltages behaves very distinctly from the predictions of the model.
The model curves tend towards zero rather quickly at fields above 1~T (blue and black lines).
Contrary, the experimental data shows abrupt deviation from the classical
calculations at these fields, saturating at approximately constant level.

From the transmission spectra in zero magnetic field and at zero gate voltage
the exact value of the refractive index  of the substrate (optical thickness) is determined experimentally~\cite{shuvaev_sst_2012}. With this parameter
the transmission in both parallel and crossed
geometries can be recalculated into the complex magneto-optical conductivity of
mercury telluride~\cite{dziom_2d_2016} without additional assumptions. The diagonal conductivity $\sigma_{xx}$
is mostly responsible for the parallel transmission in our experiments and
for the dissipation in DC transport measurements. The off-diagonal
conductivity $\sigma_{xy}$ is related to the transmission in the crossed
geometry and for the quantum Hall plateaus in the DC experiments. $\sigma_{xy}$ is especially relevant for the emergence of the universal
Faraday rotation $\alpha$ and it is plotted in the inset of
Fig.~\ref{theta_eta_7T}. The data are shown in a dimensionless form by
multiplying the conductivity $\sigma$ with the impedance of vacuum
$Z_0 \approx 377$~Ohm.

The upper inset in Fig.~\ref{theta_eta_7T} shows the imaginary
part of $\sigma_{xy}$ which appear at nonzero frequencies only. The real part of $\sigma_{xy}$ is shown in the lower inset. It also
demonstrates the deviation from the classical Drude behaviour and
saturates at the level slightly above the universal value of $Z_0 \frac{e^2}{h} =
 2 \alpha$. We attribute this deviation to the uncertainties of the experiment.

From the Drude fits of the dynamic conducivity of sample \#1 in the vicinity of cyclotron resonance the parameters of the charge carriers could be calculated which are given in Tab.~\ref{tab}. Much lower mobility of the electrons ($+500$~V) compared to holes ($-500$~V) is probably the reason that no quantized Faraday effect could be observed for positive voltages. High Dirac-hole mobility in HgTe wells can be explained by screening of their scattering by heavy holes~\cite{kozlov_jetpl_2013}.

\begin{table}
\caption{Drude parameters of the charge carriers in HgTe sample \#1 as obtained from the fits of magneto-optical conductivity: density $n_{2D}$, effective mass $m/m_0$, and mobility $\mu$. The gate voltage~$  - 500$~V corresponds to hole carriers, and~$  + 500$~V to electrons, respectively. $m_0$ is the free electron mass.} \label{tab}
\begin{tabular}{cccc}
\hline
Gate (V) &  $n_{2D}$(cm$^{-2}$) & $m/m_0$  & $\mu$~(cm$^2/$Vs) \\
\hline \hline
-500~V & $(3.3\pm0.5)\times 10^{10}$ & $(7.5 \pm 1)\times 10^{-3}$ & $(6.6 \pm 1.0)\times 10^{4}$ \\
\hline
+500~V & $(1.4 \pm 0.3)\times 10^{11}$   & $ (9.2 \pm 1)\times 10^{-3}$
&  $(2.0 \pm 0.2)\times 10^{4}$   \\
\hline
\\
\end{tabular}
\end{table}

\begin{figure}[tbp]
\centerline{\includegraphics[width=0.95\columnwidth,clip]
{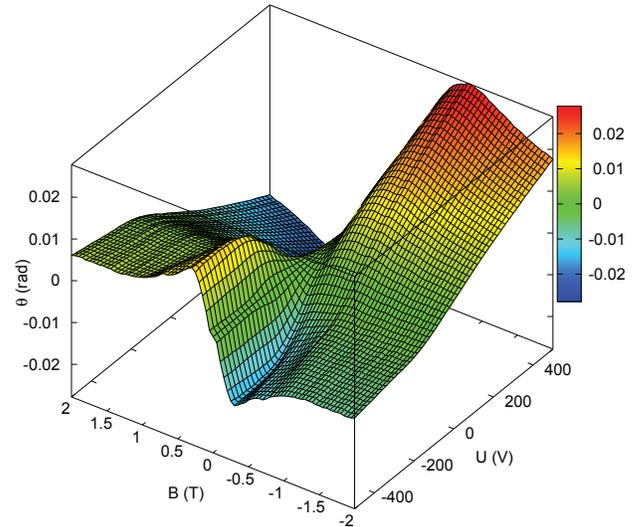}}
\caption{Faraday rotation $\theta$ of the HgTe quantum well \#1 as function of gate voltage and magnetic field. The values of $\theta$ are colour-coded for clarity. The data are given for the increasing gate voltage from -500~V to +500~V as applied to the gate electrode.
The maximum and minimum of $\theta$ at low magnetic fields are the manifestations
of the cyclotron resonance. The inversion from maximum to minimum reflects the transition from the hole-like
charge carriers at negative gate voltages to the electron-like charge carriers
at positive voltages.}
\label{efscan_3d}
\end{figure}

At a fixed frequency of the incident radiation $f = 320$~GHz and
at $T = 1.8$~K there are
two external parameters which can be tuned: magnetic field and gate voltage.
A good overview of the experimental data set obtained by changing both
parameters is provided by Fig.~\ref{efscan_3d}. Here, the color coded height
represents the Faraday rotation $\theta$ as a function of magnetic field
$-2$~T~$< B < 2$~T and of gate voltage $-500$~V~$< U < 500$~V in the
direction of increasing voltage. The data shown in Fig.~\ref{theta_eta_7T} are cuts of the parametric
surface in Fig.~\ref{efscan_3d} at fixed gate voltages. The cyclotron
resonance peaks in Fig.~\ref{theta_eta_7T} are also seen in
Fig.~\ref{efscan_3d}. However, now it is possible to see the continuous
evolution of the cyclotron resonances with the gate voltage. The positive
peak at positive magnetic fields and gate voltage of $-500$~V gradually disappears
and transforms into a negative peak at $+500$~V. This is a manifestation
of the transition from the hole-like carriers at negative gate voltages
to the electron-like carriers at the positive gates.

\begin{figure}[tbp]
\centerline{\includegraphics[width=0.95\columnwidth,clip]
{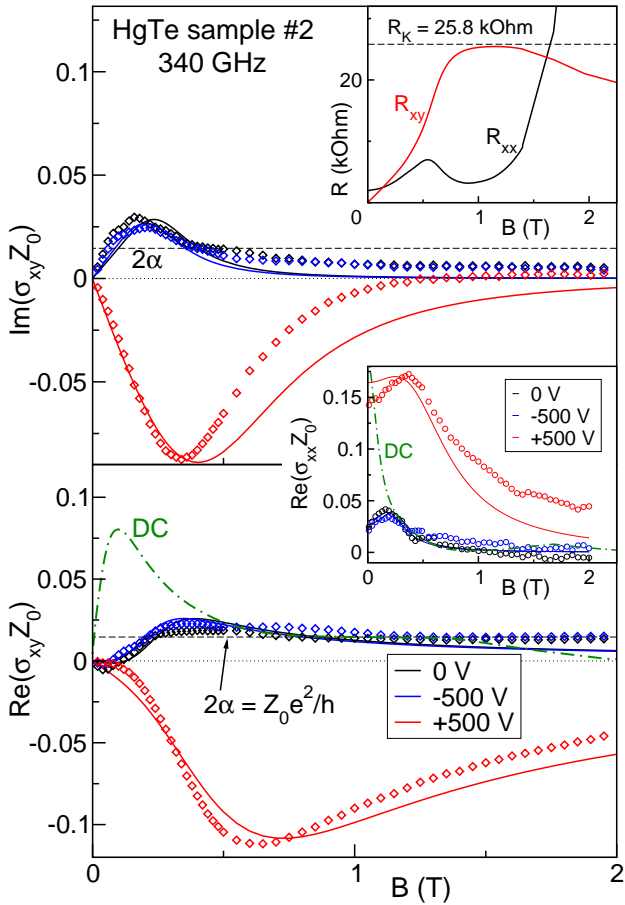}}
\caption{Magnetic field dependence of the off-diagonal conductivity
$\sigma_{xy}$ for the sample \#2 at different gate voltages.
Symbols represent experimental data,
lines are fits using the Drude model. Green dash-dotted line shows the conductivity calculated from the DC data. Black dashed lines show the universal
value of conductivity ($2 \alpha$) predicted for doubly degenerate Dirac fermions.
Upper inset shows the
DC longitudinal and Hall resistivity measured on the same sample. The divergence of $R_{\mathrm{xx}}$ in high magnetic fields is due to field-induced transition to the insulating state. Lower inset shows the diagonal dynamic conductivity $\sigma_{\mathrm{xx}}$ with Drude fit and DC data represented as solid and dash-dotted lines, respectively. The maximum in $\sigma_{\mathrm{xx}}$ close to $0.2$~T corresponds to the cyclotron resonance.}
\label{sigma}
\end{figure}

The main result of the present work, a plateau
in Faraday rotation close to the universal value $\theta=\alpha$, is supported by the measurements on sample \#2.
Eight contacts have been prepared around the edges of sample \#2, which allowed to measure DC longitudinal and
transverse resistivities $R_{\mathrm{xx}}$ and $R_{\mathrm{xy}}$. These data are shown in
the upper inset of Fig.~\ref{sigma}. The black curve is the longitudinal
resistivity $R_{\mathrm{xx}}$, the red curve represents the transverse resistivity $R_{xy}$. The pronounced plateau
at fields between 0.75 and 1.5~T is clearly seen in the
$R_{\mathrm{xy}}$ data. The value of the transverse resistivity at the plateau is
around 25.8~k$\Omega$ which can be expected if only the last quantum Hall plateau is observed and the degeneracy factor is equal to $\gamma=2$. The DC data correspond well to the universal value of
the Faraday rotation $\theta = \alpha$. Indeed, in the limit of small absorption by thin film~\cite{shuvaev_apl_2013} we may write: $\theta \sim t_c/t_p \sim t_c \sim Z_0/2R_{\mathrm{xy}}$, which leads to $\theta = \alpha$ for $R_{\mathrm{xy}}=h/e^2$. Direct correspondence between quantum Hall effect and quantized Faraday rotation is well known in ordinary 2D electron gases~\cite{volkov_jetpl_1985}.

The magneto-optical conductivity of sample \#2 is shown in
Fig.~\ref{sigma}. The imaginary part in the upper panel
reveals no plateau neither at the positive
nor at the negative gate voltages. The real part of the conductivity shown in
the lower panel demonstrates a clear plateau at fields above 1~T at zero
and negative gate voltages. The value of this plateau equals to $\sigma_{xy}Z_0=2 \alpha = Z_0 e^2/h$ and it corresponds well to the
DC data shown by green line. In the electron-like doping regime at the
positive gate voltages no such plateau is observed in magnetic fields
below 2~T.

An interesting difference between the DC and the THz
conductivity of HgTe films is that the high frequency measurements show the robust plateau in
the Faraday rotation $\theta$ up to 2~T in Fig.~\ref{sigma} and up to 7~T
in Fig.~\ref{theta_eta_7T}, whereas the DC data the plateau is limited to the field range $\sim 0.8-1.3$~T. In addition, $R_{\mathrm{xx}}$ reveal clear indication of insulating behavior above 1.5~T. This could be due to the
quantum Hall liquid-to-insulator transition~\cite{wang_prl_1994, jiang_prl_1993, shahar_prl_1995}.  In the static experiment the current
has to follow the percolation path across the whole sample and the
liquid-to-insulator transition sets in rather early. Contrary, at high frequencies
the charge carriers are able to move a small distance only ($\ell \sim \min( \upsilon_F \tau, \upsilon_F / \nu ) \sim 1 \mu m$)
during one period of the electromagnetic wave. Here~\cite{shuvaev_prb_2013} $\upsilon_F \sim 10^6 m/s$ is the Fermi velocity, $\tau \sim 10^{-12}$~s is the scattering time and $\nu$ is the frequency of the experiment. The
less conducting regions of the sample are not participating in the overall
response while the signal from the conducting parts is still present.
Therefore one can expect that the transition to insulating state will disappear at high frequencies as observed here.


In conclusion, using polarisation- and phase-sensitive terahertz transmission spectroscopy, HgTe quantum wells with critical thickness have been investigated. In  external magnetic fields a universal value of the Faraday rotation close to the fine structure constant $\theta_F= \alpha \approx 1/137$ is observed for hole-like carriers. Dynamic Hall conductivity is directly calculated from the experiment and it reveals a universal value $\sigma_{xy}=e^2/h$ which corresponds to a degeneracy $\gamma=2$ of Dirac states. The universal steps in the dynamical conductivity and Faraday angle remain robust in a broad range of external magnetic fields and gate voltages. On the electronic side of the gate voltages a classical magneto-optical behavior is observed. It can attributed to much lower mobility of negatively charged carriers.

We thank G. Tkachov, E. M. Hankiewicz, and S.-C. Zhang for valuable discussions. This work was supported by Austrian Science Funds
(I815-N16, W-1243, P27098-N27).

\bibliography{literature}

\end{document}